\documentclass[aps,showpacs,floatfix,twocolumn,byrevtex,superscriptaddress]{revtex4-1}
\usepackage{amsmath}
\usepackage{amssymb}
\usepackage{amstext}
\usepackage{amsopn}
\usepackage{amsfonts}
\usepackage{amsxtra}
\usepackage[english]{babel}
\usepackage{graphicx}
\usepackage{float}
\usepackage{bm}
\usepackage{multirow}
\usepackage{dcolumn}
\usepackage{color}
\usepackage{hyperref}

\newcommand{\icm}{\ensuremath{~\textrm{cm}^{-1}}}

\begin{document}

\title{Optical study of Dirac fermions and related phonon anomalies in antiferromagnetic compound CaFeAsF}
\author{B. Xu}
\affiliation{Center for High Pressure Science and Technology Advanced Research, Beijing 100094, China}
\affiliation{University of Fribourg, Department of Physics and Fribourg Center for Nanomaterials, Chemin du Mus\'{e}e 3, CH-1700 Fribourg, Switzerland}

\author{H. Xiao}
\email[]{hong.xiao@hpstar.ac.cn}
\affiliation{Center for High Pressure Science and Technology Advanced Research, Beijing 100094, China}

\author{B. Gao}
\affiliation{Center for High Pressure Science and Technology Advanced Research, Beijing 100094, China}

\author{Y. H. Ma}
\author{G. Mu}
\affiliation{State Key Laboratory of Functional Materials for Informatics and Shanghai Center for Superconductivity, Shanghai Institute of Microsystem and Information Technology, Chinese Academy of Sciences, Shanghai 200050, China}

\author{P. Marsik}
\author{E. Sheveleva}
\author{F. Lyzwa}
\affiliation{University of Fribourg, Department of Physics and Fribourg Center for Nanomaterials, Chemin du Mus\'{e}e 3, CH-1700 Fribourg, Switzerland}

\author{Y. M. Dai}
\affiliation{Center for Superconducting Physics and Materials, National Laboratory of Solid State Microstructures and Department of Physics, Nanjing University, Nanjing 210093, China}

\author{R. P. S. M. Lobo}
\affiliation{LPEM, ESPCI Paris, PSL University, CNRS, F-75231 Paris Cedex 5, France}
\affiliation{Sorbonne Universit\'e, CNRS, LPEM, F-75005 Paris Cedex 5, France}

\author{C. Bernhard}
\email[]{christian.bernhard@unifr.ch}
\affiliation{University of Fribourg, Department of Physics and Fribourg Center for Nanomaterials, Chemin du Mus\'{e}e 3, CH-1700 Fribourg, Switzerland}

\date{\today}
%
%

\begin{abstract}
We performed optical studies on CaFeAsF single crystals, a parent compound of the 1111-type iron-based superconductors that undergoes a structural phase transition from tetragonal to orthorhombic at  $T_s$ = 121~K and a magnetic one to a spin density wave (SDW) state at $T_N$ = 110~K. In the low temperature optical conductivity spectrum, after the subtraction of a narrow Drude peak, we observe a pronounced singularity around 300~cm$^{-1}$ that separates two regions of quasilinear conductivity. We outline that these characteristic absorption features are signatures of Dirac fermions, similar to what was previously reported for the BaFe$_2$As$_2$ system~\cite{Chen2017}. In support of this interpretation, we show that for the latter system this singular feature disappears rapidly upon electron and hole doping, as expected if it arises from a van Hove singularity in-between two Dirac cones. Finally, we show that one of the infrared-active phonon modes (the Fe-As mode at 250~cm$^{-1}$) develops a strongly asymmetric line shape in the SDW state and note that this behaviour can be explained in terms of a strong coupling with the Dirac fermions.
\end{abstract}



\maketitle

%
%
The discovery of high-temperature superconductivity in the iron-based superconductors (IBSs) has attracted extensive attention~\cite{Kamihara2008,Paglione2010}. IBSs are multiband materials with a Fermi surface that is composed of disconnected hole and electron sheets~\cite{Mazin2008,Kuroki2008,Ding2008}. This multiband structure leads to rich physics, where interband antiferromagnetic interactions are crucial for high temperature superconductivity~\cite{Mazin2008,Kuroki2008}. Contrary to the Mott insulating state of the cuprates, the parent antiferromagnetic phase of the IBSs has a metallic character~\cite{James2009} and possesses nontrivial topological properties with Dirac fermions near to the Fermi energy~\cite{Ran2009,Yin2011}. The formation of the Dirac fermion state occurs near the nodes of the spin-density-wave (SDW) gap that involves several bands and leads to complex zone folding effects~\cite{Ran2009}. The existence of a Dirac cone in BaFe$_2$As$_2$ has been first reported from angle-resolved photoemission spectroscopy and magnetic quantum oscillation measurements~\cite{Richard2010,Harrison2009}. Subsequent, theoretical and experimental studies revealed that the Dirac cones exist in pairs~\cite{Morinari2010,Huynh2011}. More recently, magneto-optical experiments have demonstrated that the Dirac fermion state in BaFe$_2$As$_2$ and SrFe$_2$As$_2$ is two-dimensional~\cite{Chen2017}. The observation of such a Dirac fermion state with protected Dirac cones in the IBSs parent compounds provides us with a new kind of topological material, after cuprates~\cite{Damascelli2003}, graphene~\cite{Zhou2006}, topological insulators~\cite{Hasan2010,Qi2011}, and Weyl semimetals~\cite{Xu2011,Wan2011PRB}. It likely inspires new research that might even unravel a connection between Dirac fermions and superconductivity.

CaFeAsF is a parent compound of the 1111 type family of the IBSs. It is isostructural to LaFeAsO and exhibits a transition from a tetragonal to an orthorhombic phase at $T_s$ = 121~K that is well separated from an antiferromagnetic transition into a SDW phase at $T_N$ = 110~K~\cite{Xiao2009,Ma2015}. This is different from the parent compound of the 122 phase for which these two transitions coincide~\cite{Rotter2008a}. Upon a partial substitution of Fe by Co, CaFeAsF becomes superconducting with a maximum $T_c$ of 22~K~\cite{Matsuishi2008}. More impressively, in rare-earth doped compounds, $T_c$ as high as 56~K can be obtained~\cite{Cheng2009}. However, the study on CaFeAsF is limited due to the small size of the single crystal and its poor quality. Recently, high-quality single crystals of CaFeAsF with a size above 1~mm were successfully grown by a self-flux method~\cite{Ma2015}. The availability of such sizable single crystals allows for an optical study of the intrinsic properties of the 1111 type IBSs, such as the nodal SDW and the resulting Dirac fermion states.

The optical conductivity probes the bulk band structure and carrier electrodynamics over a broad energy range and provides rich information on the intraband and interband transitions of possible Dirac cones. Although optical investigations into the parent compounds of IBSs have been conducted by many groups~\cite{Hu2008,Charnukha2013,Dai2016,Akrap2009,Nakajima2011,Schafgans2011,Xu2015,Xu2016,Mallett2017}, the optical signatures of the Dirac fermions have so far been only discussed for the 122 compound~\cite{Chen2017}. The evolution of these Dirac fermions with electron and hole doping has also not yet been reported. We fill this gap by performing a systematic optical study on single crystals of the CaFeAsF parent compound which reveals clear spectroscopic signatures of the Dirac fermions. Furthermore, for the BaFe$_2$As$_2$ compound, we study how these Dirac fermions evolve as a function of electron and hole doping. Finally, in CaFeAsF we observe an anomalous lineshape of the Fe-As mode which provides evidence that it strongly interacts with the Dirac fermions.

%
%
High-quality single crystals of CaFeAsF were grown using the self-flux method with a CaAs flux~\cite{Ma2015}. Their electronic and magnetic properties and the structural and antiferromagnetic transition temperatures of $T_s$ = 121~K and $T_N$ = 110~K, respectively, were reported in Ref.~\cite{Ma2015}. High-quality single crystals of BaFe$_2$As$_2$ and its electron- and hole-doped compounds [Ba(Fe$_{0.985}$Co$_{0.015}$)$_2$As$_2$ and Ba$_{0.92}$K$_{0.08}$Fe$_2$As$_2$] were synthesized with the methods described in the cited references~\cite{Shen2011,Rullier-Albenque2009}. The \emph{ab}-plane reflectivity $R(\omega)$ was measured at a near-normal angle of incidence on Bruker VERTEX 70v and IFS66v FTIR spectrometers with an \emph{in situ} gold overfilling technique~\cite{Homes1993}. Data from 50 to 12\,000\icm\ were collected at different temperatures with a ARS-Helitran crysostat. The room temperature spectrum in the near-infrared to ultraviolet range (4000 -- 50\,000\icm) was obtained with a commercial ellipsometer (Woollam VASE). The optical conductivity was obtained by performing a Kramers-Kronig analysis of $R(\omega)$~\cite{Dressel2002}. For the low frequency extrapolation below 50\icm, we used a Hagen-Rubens function ($R = 1 - A\sqrt{\omega}$). For the extrapolation on the high frequency side, we assumed a constant reflectivity up to 12.5~eV that is followed by a free-electron ($\omega^{-4}$) response.

%
%
\begin{figure}[tb]
\includegraphics[width=0.95\columnwidth]{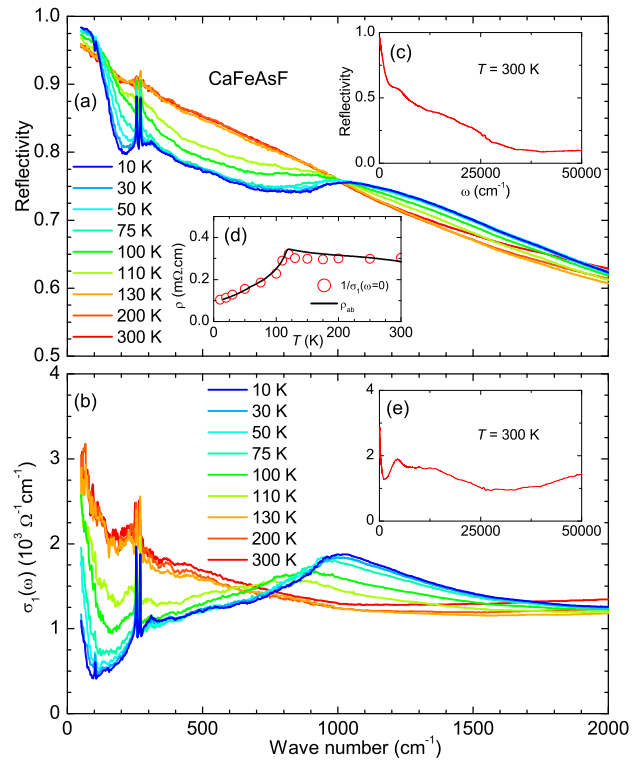}
\caption{ (color online) (a) Far-infrared reflectivity of CaFeAsF at different temperatures. Inset (c): 300-K reflectivity over a broad frequency range. Inset (d): Comparison of the $dc$ resistivity, $\rho_{ab}$ (solid line), with the zero-frequency values of the Drude fits to the conductivity data (open circles). (b) Far-infrared optical conductivity of CaFeAsF at different temperatures. Inset (e): Optical conductivity at 300~K over a broad frequency range.}
\label{Fig1}
\end{figure}
Figure~\ref{Fig1} displays the spectra of the reflectivity $R(\omega)$ [Fig.~\ref{Fig1}(a)] and the optical conductivity $\sigma_1(\omega)$ [Fig.~\ref{Fig1}(b)] of CaFeAsF up to 2\,000\icm\ at several selected temperatures. Insets~(c) and (e) show the room temperature spectra over a larger range up to 50\,000\icm. In the low-frequency region, $R(\omega)$ shows a typical metallic response, i.e. it approaches unity at low frequencies and increases upon cooling. The corresponding conductivity spectra exhibit a Drude-like peak centered at zero frequency. The metallic response is also evident from the $dc$ resistivity in inset~(d) of Fig.~\ref{Fig1}(a) as determined from a transport measurement (solid line) and from the optical conductivity (open symbols). Upon entering the SDW state, $R(\omega)$ exhibits a clear suppression between 200 and 1\,000\icm\ and an enhancement between 1\,000 and 2\,000\icm. Correspondingly, the optical conductivity $\sigma_1(\omega)$ gets suppressed below $\sim$~800\icm\ in the  SDW state and the resulting missing spectral weight is transferred to higher frequencies up to a range of about 2\,000\icm. Although the spectral weight of the Drude-like contribution is strongly reduced in the SDW state, it remains finite to the lowest temperature. The same kind of characteristic spectral features due to the formation of a SDW gap were observed in 122-type AFe$_2$As$_2$ (A = Ba, Sr and Ca) single crystals~\cite{Hu2008,Charnukha2013,Dai2016}, as well as in 1111-type  LaFeAsO~\cite{Chen2010,Dong2010}. This suggests that these undoped parent compounds have similar electronic and magnetic properties. In particular, the persistence of a finite Drude-reponse to low temperature signifies that the SDW gap develops only on parts of the Fermi-surface.

\begin{figure}[tb]
\includegraphics[width=0.95\columnwidth]{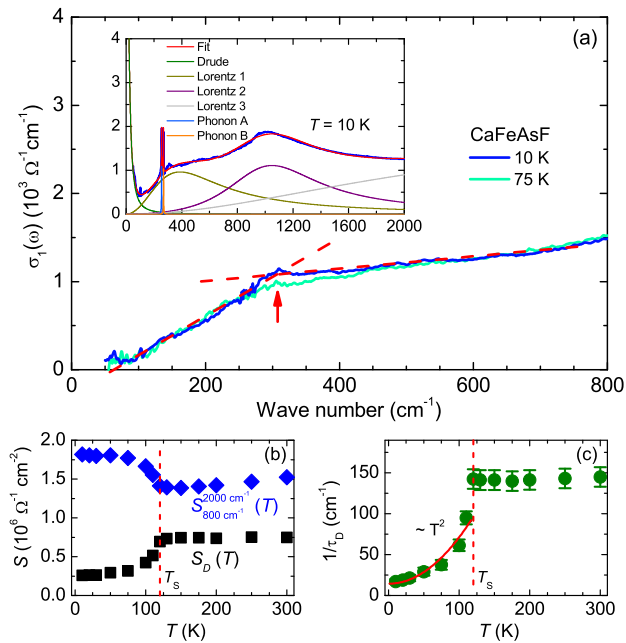}
\caption{ (color online) (a) Optical conductivity of CaFeAsF after subtraction of the Drude and phonon terms in the antiferromagnetic state. The red dashed lines through the data are linear guides to the eye. The red arrow indicates a possible van Hove singularity. Inset: Fit of $\sigma_{1}(\omega)$ at 10~K with the Drude-Lorentz model. (b) Temperature dependence of the Drude spectral weight $S_{D}(T)$ and the restricted spectral weight $S^{2000\icm}_{800\icm}(T)$. (c) Temperature dependence of the scattering rate $1/\tau_D$ of the Drude term. The solid line represents a $T^{2}$ behavior. The vertical dashed line denotes the structural transition temperature $T_s$.}
\label{Fig2}
\end{figure}
For a quantitative analysis we fitted the optical spectra with a Drude-Lorentz model,
\begin{equation}
\label{DrudeLorentz}
\sigma_{1}(\omega)=\frac{2\pi}{Z_{0}}\biggl[\frac{\Omega^{2}_{pD}}{\omega^{2}\tau_D + \frac{1}{\tau_D}} + \sum_{k}\frac{\gamma_{k}\omega^{2}S^{2}_{k}}{(\omega^{2}_{0,k} - \omega^{2})^{2} + \gamma^{2}_{k}\omega^{2}}\biggr],
\end{equation}
where $Z_{0}$ is the vacuum impedance. The first term describes the free-carrier Drude response, characterized by a plasma frequency $\Omega_{pD}$ and a scattering rate $1/\tau_D$. The squared frequency, $\Omega^2_{pD} \propto n/m$, is a measure of the ratio of the electron density, $n$, and their effective mass, $m$. The second term contains a sum of Lorentz oscillators for which each has a resonance frequency $\omega_{0,k}$, a line width $\gamma_k$ and an oscillator strength $S_k$. The corresponding fit to the conductivity at 10~K (thick blue line) using the function of Eq.~\ref{DrudeLorentz} (thick red line) is shown in the inset of  Fig.~\ref{Fig2}(a). As shown by the thin coloured lines, the fitting curve is decomposed into one Drude term representing the response of free carriers, two Lorentz terms depicting the broader electronic peaks around 300 and 1\,000\icm, two Lorentz terms for the two sharp phonons, as well as a Lorentz term that accounts for the interband transitions at higher energy. With this fit configuration we could also describe with equal quality the $\sigma_1(\omega)$ curves at all temperatures.

Fig.~\ref{Fig2}(b) shows the temperature dependence of the Drude weight, $S_D(T) = \frac{\pi^2}{Z_0}\Omega^2_{pD}$, obtained from these fits with Eq.~\ref{DrudeLorentz}. Above $T_s$, $S_D(T)$ is almost temperature independent and reflects the weak temperature dependence of $\sigma_{1}(\omega)$. Toward lower temperatures there is a strong decrease of the Drude weight due to the opening of the SDW gap. In agreement with a nodal nature of the SDW gap, the Drude weight is strongly suppressed but remains finite at low temperature. Notably, the onset of this decrease is rather sharp and occurs around $T_s$ (marked by the dashed red line) rather than at $T_N$. This result demonstrates that the electronic state near the structural phase transition is already very close to that in the magnetic state, which may imply an intimate relationship between the structural and magnetic transitions~\cite{Ma2008,Fang2008,Xu2008}. A corresponding effect is seen in Fig~\ref{Fig2}(c) for the scattering rate, $1/\tau_D$, which also shows a strong reduction that sets in below $T_s$. This strong reduction of the scattering rate can also be explained in terms of the Dirac-cone-like band dispersion at the nodes of the SDW gap~\cite{Ran2009,Richard2010}. The red solid line denotes a $T^2$ fit of $1/\tau_D$ that spans the entire temperature range below $T_s$ and is suggestive of a Fermi-liquid behavior.

Fig.~\ref{Fig2}(a) shows the conductivity spectra at 10 and 75~K, after the contribution of the Drude response and the sharp phonon modes have been subtracted to emphasize the singular structure of the low energy absorption peak. For the 10~K spectrum it clearly reveals a singular kink around 300\icm\ and two quasilinear regions toward lower and higher frequency shown by the red dashed lines. The spectrum at 75~K confirms that with the temperature increasing toward $T_s$ this kink weakens and becomes less singular. It also shows that the linearly increasing part of the conductivity at lower and higher frequency is hardly temperature dependent. This confirms that the missing spectral weight of the Drude-peak is transferred to the broad band around 1000\icm\ that corresponds to the SDW gap. This assignment is supported by the temperature dependence of the restricted spectral weight, defined as
$S(T) = \int_{\omega_a}^{\omega_b}\sigma_{1}(\omega,T)d\omega$, with the cutoff frequencies $\omega_a$ = 800\icm\ and $\omega_b$ = 2\,000\icm, that is shown by the blue diamonds in Fig.~\ref{Fig2}(b).

Next, we discuss in more detail the origin of the singular absorption feature around 300\icm. In most of the previous studies of BaFe$_2$As$_2$ it was attributed to a second, smaller SDW gap~\cite{Hu2008,Akrap2009,Nakajima2011,Charnukha2013,Dai2016}. However, a real challenge for this interpretation has been provided by the optical spectra of a detwinned BaFe$_2$As$_2$ single crystal for which the absorption peak in the response along the $b$-axis acquires a very singular shape~\cite{Nakajima2011}. More recently, the magnetic-field dependence of this absorption feature has been interpreted in terms of an interband transition of two-dimensional Dirac fermions~\cite{Chen2017} that appear in the nodal regions of the SDW gap due to band backfolding effects. Here, we follow up on this interpretation for which the Dirac nodes are protected by the combination of the physical symmetry and the topology of the band structure~\cite{Ran2009,Richard2010,Harrison2009}. The linear slope of the conductivity in Fig.~\ref{Fig2}(a) is typical for the linear band dispersion of Dirac fermions~\cite{Hosur2012PRL,Timusk2013PRB}. The additional singular kink around 300\icm\ is characteristic of the single-point singularity in Weyl semimetals, where a van Hove singularity due to saddle points of the conduction and valance bands arises between a pair of Dirac cones with opposite chiralities~\cite{Tabert2016PRB,Xu2016PRB,Kimura2017}. In the pnictides case, theory also predicts a pair of Dirac cones but with the same chirality~\cite{Morinari2010}. These similarities between CaFeAsF and Dirac fermion materials suggest that the observed features are related to the contributions of Dirac fermions. Meanwhile, the presence of the finite Drude response at low frequencies, representing the intraband transition contributions of the Dirac cones, suggests a non-zero chemical potential (relative to the Fermi level) of the Dirac nodes, which would result in a redistribution of the low-energy spectral weight between intraband and interband transitions. In addition, we should notice that in CaFeAsF $\sigma_{1}(\omega)$ is almost frequency independent in the range between 300 and 600\icm\ if the contributions of other interband transitions at higher frequencies are subtracted. Such plateau-like behavior of $\sigma_{1}(\omega)$ is indicative of the optical response of two-dimensional Dirac cones~\cite{Chen2017,Terashima2018,Sugimoto2011}, which is similar to what has been predicted and observed in graphene~\cite{Mak2008,Kuzmenko2008}. Very recently, the existence of Dirac fermions in CaFeAsF has been confirmed by quantum oscillation measurements~\cite{Terashima2018}.

\begin{figure}[tb]
\includegraphics[width=0.95\columnwidth]{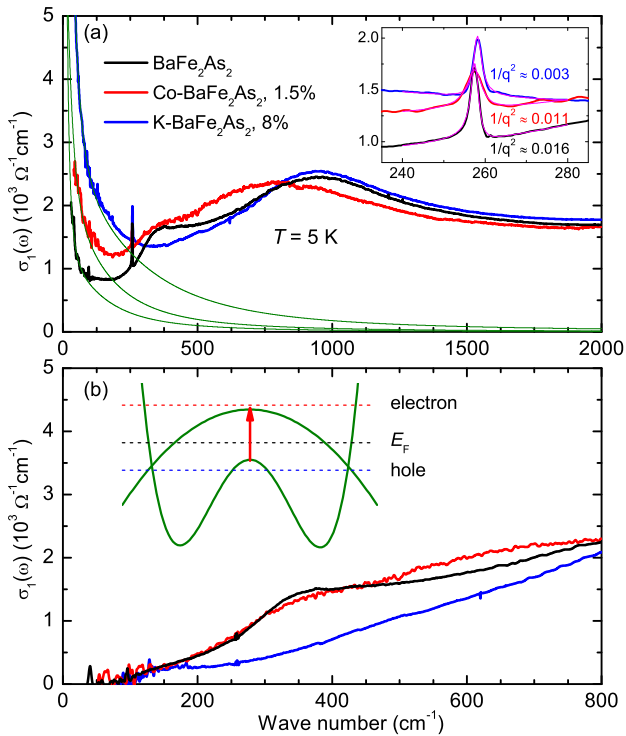}
\caption{ (color online) (a) Optical conductivity of undoped, Co (electron) and K (hole) doped BaFe$_2$As$_2$ at 5~K. The inset shows the infrared-active phonon and the corresponding Fano fits. (b) The subtracted spectra in the low-energy region. The diagram shows a schematic band dispersion of a pair of Dirac cones with the same chirality~\cite{Morinari2010}.}
\label{Fig3}
\end{figure}
This connection between the singular features in the low-frequency part of the optical response and the Dirac nodes in the parent compounds of the IBSs has been further tested by studying how they evolve as a function of electron and hole doping. Figure~\ref{Fig3} shows the corresponding $\sigma_1(\omega)$ spectra (at 5 K) for the parent compound BaFe$_2$As$_2$ (black line) as well as for weakly electron doped Ba(Fe,Co)$_2$As$_2$, (red line) and weakly hole doped (Ba,K)Fe$_2$As$_2$ (blue line). For both cases the kink feature can be seen to diminish very rapidly. A very weak feature is still visible at the same energy for the more weakly doped Ba(Fe,Co)$_2$As$_2$. This can be readily understood in terms of the sketch of the band structure in the vicinity of a pair of Dirac cones carrying the same chirality that is shown in Fig.~\ref{Fig3}(b) and describes the expected behaviour in the antiferromagnetic state of an IBSs parent compound~\cite{Morinari2010}. The red arrow indicates the interband transitions between the two saddle points in the middle of the pair of Dirac cones, which give rise to the singular kink in $\sigma_1(\omega)$. This singular feature exists only as long as the chemical potential ($E_F$) is located between the two saddle points. Thus it diminishes rather rapidly as $E_F$ is shifted upwards (downwards) upon electron (hole) doping and vanishes as soon as $E_F$  is located outside the energy window between these two saddle points (red and blue dotted lines). The position of $E_F$ with respect to these saddle points can also change for different parent compounds such as for the family of AFe$_2$As$_2$ with A = Ba, Sr, and Ca~\cite{Hu2008,Charnukha2013,Dai2016}. The reported optical data show indeed that this singularity is more pronounced for Ba than for Sr and essentially absent for CaFe$_2$As$_2$~\cite{Hu2008,Charnukha2013,Dai2016}. The latter behavior of CaFe$_2$As$_2$ agrees with a theoretical calculation which predicts that $E_F$ is located outside the energy window of the saddle points~\cite{Morinari2010}.

\begin{figure}[tb]
\includegraphics[width=0.95\columnwidth]{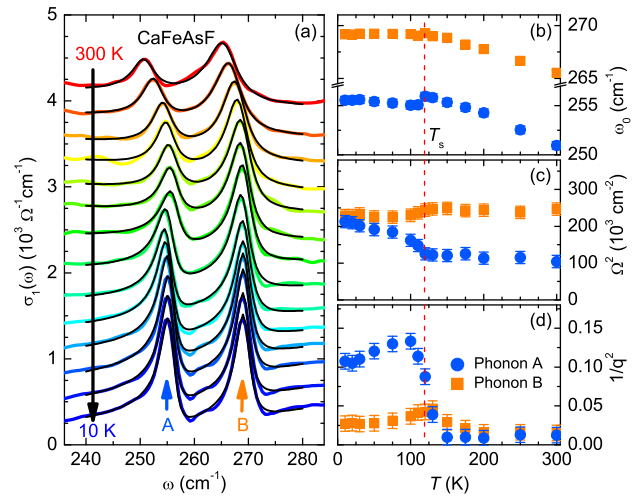}
\caption{ (color online) (a) Line shape of the infrared-active phonon modes at temperature from 300 to 10~K. The black solid lines through the data denote the Fano fits. (b--d) Temperature dependence of the resonance position $\omega_0$, intensity $\Omega^2$ and Fano parameter $1/q^2$ of the two phonons. The vertical dashed line denotes the structural transition temperature $T_s$.}
\label{Fig4}
\end{figure}
Next, we turn to the anomalous temperature dependence of the lineshape of the phonon modes across the antiferromagnetic transition in CaFeAsF. As shown in Fig.~\ref{Fig1}, in addition to the gross features, two sharp peaks representing the symmetry allowed in-plane infrared active $E_u$ phonon modes are observed around 250 and 265\icm. The mode at lower frequency (phonon A) is assigned to the in-plane displacement of the Fe-As ions, which is similar to the phonon features observed at the same energy in 122-type AFe$_2$As$_2$~\cite{Akrap2009,Hu2008,Charnukha2013,Dai2016} and 1111-type ReFeAsO (Re = La, Nd and Sm)~\cite{Dong2010}. The other mode (phonon B), which is located at significantly higher frequency in ReFeAsO~\cite{Dong2010}, is associated with the in-plane displacement of the F-Ca ions. Figure~\ref{Fig4}(a) shows an enlarged view of $\sigma_{1}(\omega)$ in the relevant frequency region of 235--285\icm, where the two phonon modes are seen more clearly and several interesting features can be identified. Firstly, the position of phonon A shifts discontinuously at $T_s$, while phonon B does not show a corresponding renormalization across the structural and magnetic transitions. Secondly, the line shapes for both phonon modes become rather asymmetric toward low temperatures. Such an asymmetric line shape is characteristic of a strong coupling between the phonons and some electronic excitations.

The asymmetry of the phonon has been quantified by fitting with a so-called Fano-model~\cite{Fano1961PR} that can be written as
\begin{equation}
\label{Fano}
\sigma_{1}(\omega)=\frac{2\pi}{Z_0} \frac{\Omega^2}{\gamma}
\frac{q^2 +\frac{4q (\omega - \omega_0)}{\gamma} -1}{q^2 (1 + \frac{4(\omega - \omega_0)^2}{\gamma^2})},
\end{equation}
where $\omega_{0}$, $\gamma$ and $\Omega$ represent the resonance frequency, line width and strength of the phonon, respectively. The asymmetry of this Fano-type line shape is described by the dimensionless parameter $1/q^2$. As $1/q^2$ increases, the line shape becomes more asymmetric, indicating a larger coupling strength. In the case of $1/q^2 = 0$, a symmetric Lorentz-type line shape is recovered.

Figure~\ref{Fig4}(a) shows the corresponding fits of the phonon line shapes at different temperatures (black solid lines) which describe the measured spectra rather well. Figures~\ref{Fig4}(b--d) show the temperature dependence of the fitting parameters $\omega_{0}$, $\Omega^2$ and $1/q^2$. Clearly, phonon A is much more strongly renormalized across the structural and magnetic transitions than phonon B. It exhibits pronounced anomalies in all three fitting parameter similar to the case of BaFe$_2$As$_2$~\cite{Akrap2009,Schafgans2011}. Nevertheless, the low temperature value of $1/q^2 \simeq 0.1$ of phonon A is much larger than the one for BaFe$_2$As$_2$ ($1/q^2 \simeq 0.016$)~\cite{Schafgans2011}. This corresponds to a much larger asymmetry of the phonon line shape and thus a larger coupling strength between the phonon and the electronic excitations in CaFeAsF. The much weaker normalization of the phonon B suggests that the lattice vibrations associated with the F-Ca ions are less sensitive to the structural and magnetic transitions and the related changes of the electronic structure.

It should be noted that the Dirac fermions may be responsible for the enhanced electron-phonon coupling in CaFeAsF. In Dirac materials, the Dirac nodes are located in close proximity to the Fermi level, and the corresponding excitations occur at very low energy. When this energy scale overlaps with the phonon resonance, a strong coupling between the Dirac fermions and phonon can arise and manifest itself in a strong Fano resonance~\cite{Sim2015,LaForge2010,Xu2017}. Indeed, in CaFeAsF the low-energy absorption associated with the Dirac fermions, especially the enhanced singular absorption around 300\icm, is rather close the phonon resonance at 255\icm. The weaker electron phonon coupling in BaFe$_2$As$_2$ thus can be explained in terms of a small blue shift of the singular absorption feature which occurs, here, around 350\icm. Furthermore, the inset of Fig.~\ref{Fig3}(a) shows the evolution of the asymmetry of the phonon mode in Ba(Fe,Co)$_2$As$_2$ and (Ba,K)Fe$_2$As$_2$, which confirms that its strength is correlated with the one of the singular absorption feature.

Finally, we discuss the possible anisotropic optical responses of the Dirac fermions in the parent compounds. A recent theoretical study has revealed that the optical response of the Fermi arc surface states of a Dirac semimetal can be strongly anisotropic with respect to the polarization direction of the light, i.e. whether it is polarized along or  transverse to the Fermi arc~\cite{Shi2017}. This may also explain the pronounced anisotropy of the optical response that has been observed in polarized infrared studies of detwinned BaFe$_2$As$_2$ crystals~\cite{Nakajima2011}. There the singularity in the electronic part and the anomaly of the Fe-As phonon mode become much more pronounced along the $b$-axis than along the $a$-axis direction. In analogy, this can be understood in terms of an anisotropic light-matter interaction which depends on whether the light polarization is parallel or perpendicular to the line connecting the pair of Dirac nodes. This calls for further polarized infrared studies on detwinned CaFeAsF for which this kind of electron phonon coupling effects should be even stronger since the Fe-As phonon mode is closer to the electronic singularity.

%
%
To summarize, the optical conductivity of CaFeAsF single crystal has been measured as a function of temperature and over a broad frequency range. We have studied how the spectral features associated with the SDW gap develop across the structural and magnetic transitions. In the SDW state we have observed a singular absorption feature and two regions of quasilinear conductivity that have been interpreted in terms of the response of Dirac fermions. We also showed that these features disappear very rapidly as a function of electron and hole doping in the BaFe$_2$As$_2$ system. In CaFeAsF we have furthermore found that the infrared-active Fe-As phonon mode shows signatures of a strong coupling with the Dirac fermions that appear in the SDW state. The strength of this electron phonon coupling can be material dependent, since it strongly depends on the relative position of the electronic singularity and the phonon mode, and can also be strongly anisotropic with respect to the polarization of the light, i.e. whether it is parallel or perpendicular to the line that connects the pair of Dirac cones.

%
%
Work at HPSTAR was supported by NSAF, Grant No. U1530402. Work at the University of Fribourg was supported by the Schweizer Nationalfonds (SNF) by Grant No. 200020-172611.

%

\end{document}